# FRA-DiagSys: A Transformer Winding Fault Diagnosis System for Identifying Fault Types and degrees Using Frequency Response Analysis

Guohao Wang[1]

Guohao Wang: guohaowang@ieee.org;

**Abstract:** The electric power transformer is a critical component in electrical distribution networks, and the diagnosis of faults in transformers is an important research area. Frequency Response Analysis (FRA) methods are widely used for analyzing winding faults in transformers, particularly in Chinese power stations. However, the current approach relies on manual expertise to interpret FRA curves, which can be both skill-intensive and lacks precision. This study presents a novel approach using a Multilayer perceptron model to directly model and analyze FRA data, simulating various winding fault types and degrees in 12-disc winding and 10-disc winding transformers with different connection configurations, resulting in three distinct datasets. Six different Multilayer perceptron architectures were developed, with optimal models achieving recognition accuracies of over 99.7% for diagnosing fault degrees and more than 90% for fault types. Hence, this paper has yielded a model architecture that exhibits commendable performance in diagnosing various fault types and their severities in different models of transformers when utilizing different FRA connection methods. Additionally, a specialized diagnostic system called FRA-DiagSys with two-stage model utilization was developed, achieving 100% accuracy in diagnosing fault types and degrees for a specific winding-10 power transformer, surpassing other diagnostic methods and strategies.

**Highlights:**

1. Developed deep learning models utilizing frequence response data for winding fault diagnosis.
2. Validated the model's superior performance on three laboratory windings datasets.
3. Proposed a two-stage winding detection and diagnosis system with an accuracy of 100%.
4. Statistically demonstrated the impact of CIW and EE wiring configurations on fault diagnosis.

# 1. Introduction

With the advancement of power systems, power devices such as power transformers have become ubiquitous due to their voltage conversion capabilities. Power transformers play a pivotal role in electrical systems, but they come with high costs and the potential for significant faults. Hence, there is a pressing need for a precise fault diagnosis system to carry out transformer fault detection tasks. In recent years, with the development of hardware instruments, a series of hardware-based detection methods have emerged. Among these methods, FRA (*IEEE SA - IEEE C57.149-2012*, n.d.) have proven to be an effective and time-efficient approach for extracting precise information about the operational status of transformers. It assesses a transformer's condition by analyzing its response to varying frequencies of applied voltage or current. FRA generates a curve representing impedance magnitude over a range of frequencies. Deviations from a healthy baseline curve can reveal internal faults or structural changes, aiding in early fault detection and maintenance. FRA's sensitivity and ability to detect incipient issues make it an essential method for ensuring the reliability and longevity of power transformers.

FRA devices capture the frequency response of power transformers, typically presented graphically for manual visual analysis. While there have been numerous studies (Ludwikowski et al., 2012; Mitchell & Welsh, 2017; Picher et al., 2017; Pourhossein et al., 2012; Secue & Mombello, 2008; Wang et al., 2018) attempting to explain the relationship between FRA curves and winding faults, aiming to provide empirical guidance for manual visual analysis of FRA curves, this primitive approach still presents several limitations. Firstly, although, many prior studies have employed statistical learning methods for auxiliary analysis of FRA curves (Behjat & Mahvi, 2015; Bigdeli & Abu-Siada, 2022; Parkash & Abbasi, 2023), they have not established models capable of replacing manual expertise in diagnosing transformer winding faults. Hence, experienced workers are still required for such fault diagnosis. Secondly, FRA curve distributions of transformer windings vary under different models and operating conditions. The lack of quantitative representation in manual expertise makes it challenging to transfer knowledge of FRA curves, resulting in a scenario where a worker can only analyze FRA curves of a few common transformer windings they frequently encounter. Lastly, due to the absence of quantitative analytical methods, solely relying on manual curve analysis for diagnosing different fault types and fault severities cannot achieve a high level of accuracy. Thus, there is an urgent need to develop an expert diagnostic system tailored to specific power transformer units.

Recently, with the advancement of computing resources and the growth of machine learning techniques, a plethora of classifiers have emerged. Previous works (Akhavanhejazi et al., 2011; Bigdeli et al., 2012, 2021; Mao et al., 2020; Zhao et al., 2017) utilized statistical learning approaches with human feature extraction in combination with Support Vector Machines (SVM), or decision tree (DT) classifiers for FRA-based transformer fault diagnosis. Other works incorporated FRA image information and undirect interpretations (Duan et al., 2021; Moradzadeh et al., 2022; Zhao et al., 2019; Zhou et al., 2020) for diagnostic analysis. However, these approaches did not directly utilize the numerical input data from FRA for model training, resulting in suboptimal data utilization and accuracy. In order to improve model accuracy, various numerical indices (Samimi & Tenbohlen, 2017; Zhao et al., 2021) and feature selection methods (Tahir & Tenbohlen, 2021) were proposed. While these models demonstrated high accuracy for specific units, they suffered from the drawbacks of time-consuming manual feature extraction and overfitting to specific power transformer units due to limited data sources and volumes, thereby lacking robustness and generalizability. Therefore, in this article, we propose a novel approach, not previously reported, for directly utilizing numerical data from the FRA curve in model construction. Due to the enhanced utilization of raw FRA data afforded by this approach, the model demonstrates a remarkably high level of accuracy across various diagnostic tasks, and achieved better generalizability on different dataset in this paper.

Furthermore, FRA datasets from power transformer are scarce, and real electrical power transformers are expensive to manufacture, making it challenging to create authentic datasets that encompass multiple fault types. Addressing the issue of limited data for power transformers, researchers proposed different segmentation approaches based on frequency slicing for training (Bigdeli et al., 2021; Zhao et al., 2019). However, this method still fails to fully exploit the information obtained from the entire frequency range of FRA instruments, and the testing results did not establish its broad applicability. Additionally, the reduced feature dimensionality in the training data resulted in decreased model accuracy. Therefore, in this study, simulated windings were employed to construct the dataset. Through the use of these winding templates in experiments, a comprehensive FRA dataset containing various fault types was acquired.

Meanwhile, deep learning models have made significant strides in recent years, benefiting from advances in computational power. Neural network-based model, such as extreme learning machine (ELM) (Huang et al., 2006) has shown its ability to finish the task of fault diagnosis (Ahila et al., 2015; Wong et al., 2014), exemplified by Multilayer Perceptron, Multi-layer models, as opposed to single-layer ones (ELM), can better capture the deep-level information in FRA responses and exhibit improved robustness (LeCun et al., 2015). In the past decade, within the realm of image processing, models such as Convolutional Neural Networks (CNN) (Krizhevsky et al., 2012) and ResNet (He et al., 2016), as well as in Natural Language Processing (NLP), architectures like Recurrent Neural Networks (RNN) (Sutskever et al., 2014) and multi-head attention mechanisms (Vaswani et al., 2017), have demonstrated notable performance across various domains. Regardless of whether one considers MLP or these aforementioned models, they all share a common foundation based on matrix multiplication complemented by activation functions, with model training conducted through gradient-based backpropagation. However, when dealing with the one-dimensional FRA data format, these architectures, originally tailored for their respective domains of images and text, face inherent dissimilarities due to the distinct nature of FRA data. Moreover, these models, in comparison to MLP, exhibit larger parameter counts and scale, often necessitating training and deployment on Graphics Processing Unit (GPU). Yet, in practical transformer operational environments, the available resources are frequently limited to outdated personal computers equipped solely with Central Processing Unit (CPU). Consequently, this paper elects to adopt a Multilayer Perceptron as the foundational neural network architecture. The primary objective is to investigate a set of MLP-based models, with the aim of developing a diagnostic system for transformer winding FRA data that offers high accuracy and is amenable for deployment in real-world scenarios. Hence, this paper outlines the design and evaluation of six MLP-based models for the purpose of architectural research, ultimately culminating in the proposal of a methodology for constructing a diagnostic system.

The workflow diagram of this article is illustrated in Figure 1. In light of these considerations, this paper proposes MLP-based models capable of diagnosing various faults in power transformers and assessing the degree of faults for different fault types. To demonstrate the robustness of this model architecture, we fabricated physical models of multiple transformer windings with various faults. Experimental methods were employed to obtain FRA curves for multiple power transformers under various fault conditions. Using these diverse datasets, the model was trained and tested, yielding excellent performance across various evaluation metrics. Lastly, compared to the ensemble learning results, a two-stage model utilization approach is employed to develop a transformer diagnostic system (FRA-DiagSys), which, specifically for the 10-disc power transformer winding model discussed in this paper, achieves a diagnostic accuracy of 100%.

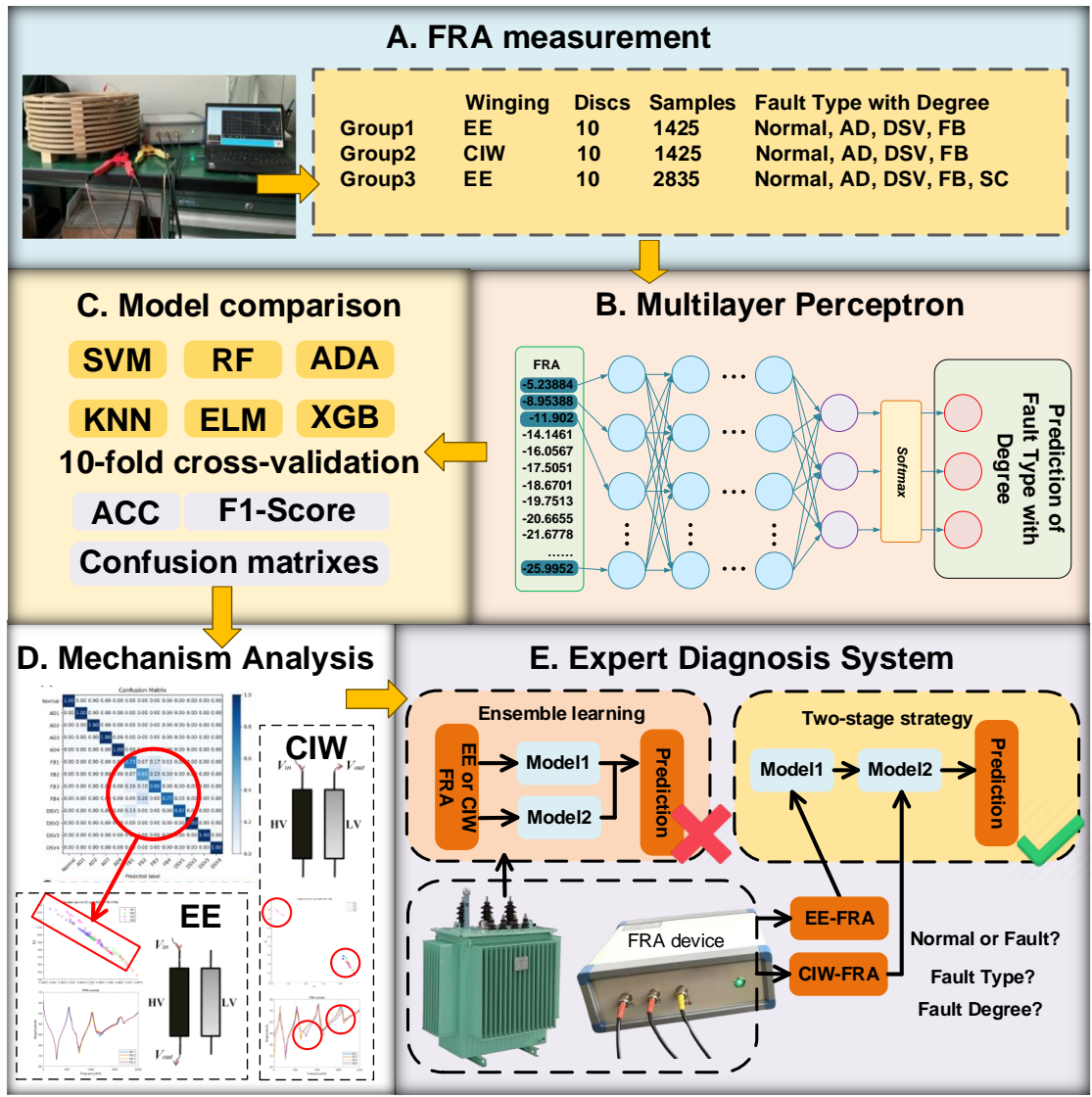

**Figure 1.** Workflow of this paper. (A) FRA measurement and dataset formation. (B) FRA-MLP model. (C) Model evaluations and comparations. (D) Mechanism Analysis of different winding methods. (E) Expert diagnosis system with two-stage method for Transformer winding.

## 2. Method

## 2.1. Data acquisition

Given the inherent complexities and cost constraints associated with introducing diverse winding fault configurations directly onto operational transformers, this study employed a meticulously crafted laboratory winding model characterized by a disc-type design. This model was used to facilitate FRA tests, enabling the acquisition of datasets essential for subsequent quantitative diagnostic analyses. The experimental configuration, as well as the specific details regarding the FRA measurement setup and the winding model, are graphically represented in Figure 1 (A).

In order to demonstrate the diagnostic accuracy of our model across various fault types in different connection configurations of transformers with varying winding designs, three distinct sets of experiments were meticulously designed to generate datasets. These experiments entailed the creation of winding models for both a 10-disc and a 12-disc transformer, involving the configuration of EE (End-to-End voltage ratio) and CIW (Capacitive Inter-winding) connection modes for the 10-disc transformer model, encompassing normal winding conditions as well as instances of varying degrees of winding faults, specifically, axial dis-placement (AD), disc space variation (DSV), and free buckling (FB). For the 12-disc transformer model with EE connections, the winding configurations included normal winding conditions, as well as winding faults such as AD, DSV, FB, and short-circuits (SC). By conducting tests on transformer windings under these different connection modes and fault scenarios, three distinct FRA datasets, namely Group1, Group2, and Group3, were ultimately generated. By conducting tests on normal samples and various fault samples with different degrees of fault at different time points, we obtained over a thousand samples. The quantities of normal samples and various fault samples in the three datasets are presented in Table 1.

Table 1. The number of various fault types in the three datasets.

|        | Normal | AD  | DSV | FB   | SC  | SUM  |
|--------|--------|-----|-----|------|-----|------|
| Group1 | 25     | 200 | 600 | 600  | 0   | 1425 |
| Group2 | 25     | 200 | 600 | 600  | 0   | 1425 |
| Group3 | 45     | 200 | 855 | 1080 | 675 | 2835 |

### 2.1.1. Creating simulated windings of power transformers

To validate the applicability of the fault detection model on transformer windings with varying numbers of discs, we fabricated two sets of simulated double-column transformers, one comprising 10 discs and the other comprising 12 discs. In order to compare the accuracy of the model in diagnosing faults based on FRA data obtained through different wiring configurations, we conducted measurements on the 10-disc winding transformer using both the EE and CIW wiring configurations. The detailed characteristics of the simulated windings for the 10-disc winding transformer and 12-disc winding transformer are presented in Table 2.

**Table 2.** The parameter values of the 10-disc winding transformer and 12-disc winding transformers.

| Winding | Parameter | 10-disc winding | 12-disc winding |
|---|---|---|---|
| HV | outer diameter | 934mm | 467mm |
| | inner diameter | 780mm | 390mm |
| | height | 205mm | 205mm |
| | number of discs | 10 | 12 |
| LV | outer diameter | 682mm | 341mm |
| | inner diameter | 520mm | 260mm |
| | height | 205mm | 205mm |
| | number of discs | 10 | 12 |

**Note:** 'HV' represents the winding parameters of the high-voltage side, while 'LV' denotes the winding parameters of the low-voltage side.

### 2.1.2. Fault simulation of the power transformer winding

This article individually simulates four types of faults, namely, free buckling (FB), axial displacement (AD), disc space variation (DSV) faults and short-circuits (SC). The details are as follows:

1) Axial displacement (AD) fault

AD fault is the vertical displacement of the winding from its original position. In this study, this fault with different degrees is realized by adding different heights of spacers at the bottom of the HV winding, as depicted in Figure 2. (A). The relationship between fault degrees and the displacement variation of the HV winding is described in Table 3.

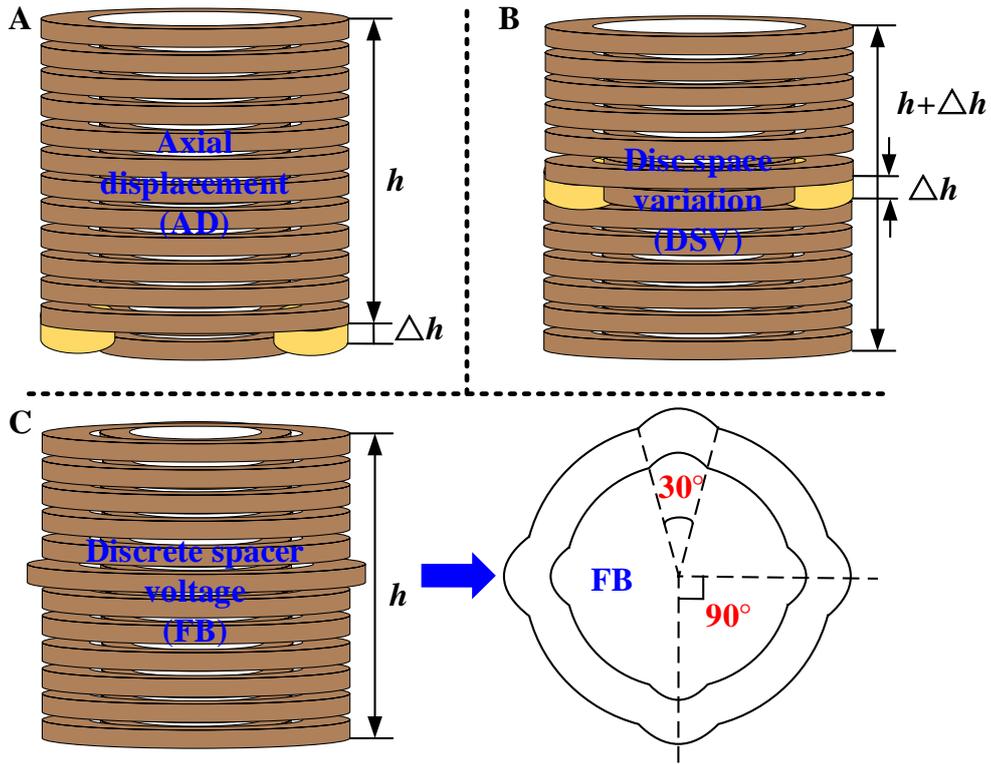

**Figure 2.** Schematics of different winding faults. (A) AD fault schematic diagram. (B) Schematic representations of FB fault. (C) Schematic diagram of DSV fault.

**Table 3.** Relationship between the AD fault degree and the displacement variation.

| Fault degree | Displacement variation /mm | |
|---|---|---|
| | 10-disc winding | 12-disc winding |
| AD1 | 10 | 10 |
| AD2 | 15 | 20 |
| AD3 | 20 | 30 |
| AD4 | 25 | 40 |

2) Free buckling (FB) fault

In a two-winding transformer, the outer winding experiences tensile tangential stresses in the outward direction due to the interplay between axial leakage flux and circumferential winding current. The occurrence of a Fault in the Windings due to excessive tensile tangential forces (referred to as an 'FB fault') becomes probable when the material constituting the conductor surpasses its elastic limit. This research endeavors to simulate a radial FB fault within the High Voltage (HV) winding, as delineated in Figure 2. (C). A solitary fault disc serves as the reference standard for fault severity. By augmenting the quantity of fault discs, four distinct levels of fault severity are established. Moreover, a systematic substitution of four levels of fault discs with non-deformable discs in the HV winding is executed. It is noteworthy that each individual protrusion on the disc exhibits an angular span of approximately 30 degrees, with the maximum radial extent of the protruding segment amounting to roughly 12.5% of the radius, denoted as $\Delta r/r=0.125$.

3) Disc space variation (DSV) fault

In contrast to the Anomalous Displacement (AD) fault, the Discrete Spacer Voltage (DSV) faults are induced by the introduction of spacers at three distinct locations along the High Voltage (HV) winding, as elucidated in Figure 2. (B). The establishment of four distinct fault degrees is accomplished by configuring four discrete spacing intervals, as detailed in Table 4. And in this paper, an equivalent spacing interval employed at different positions signifies a uniform fault degree.

Table 4. Relationship between DSV fault degree and spacing distance.

| Fault degree | Spacing distance /mm | |
|---|---|---|
| | 10-disc winding | 12-disc winding |
| **DSV1** | 5 | 10 |
| **DSV2** | 10 | 20 |
| **DSV3** | 15 | 30 |
| **DSV4** | 20 | 40 |

4) Short-circuits (SC) fault

The insulating materials between the windings of a transformer disc are susceptible to electrical forces generated by short-circuit impacts, leading to insulation damage, which in turn results in inter-turn short circuits. These short circuits, when subjected to high temperatures, can further deteriorate the insulation, potentially triggering more severe short-circuit incidents. In this study, we employed an approach involving the artificial short-circuiting of adjacent disc sections within the winding model to simulate inter-turn short-circuit faults. Four distinct fault severities were established, involving the short-circuiting of 1 to 4 disc sections, thereby simulating varying levels of short-circuit severity. Notably, a greater number of disc sections being short-circuited corresponded to higher levels of short-circuit severity.

*2.2. Deep learning models*

*2.2.1. Multilayer Perceptron*

Multilayer Perceptron (MLP) (Rumelhart et al., 1986) is a type of feedforward neural network that comprises an input layer, multiple hidden layers, and an output layer, where each layer is fully connected. The term "fully connected" implies that all nodes within a particular layer are connected to and weigh all nodes in the preceding layer within the MLP. During the training process, the MLP employs the backpropagation technique. Due to its Multiple layers, MLP qualifies as a deep learning technique. For the three FRA datasets, as each sample in the dataset consists of 2000 discrete points, the input is represented as a single vector of length 2000. denoting as vector $x$, the computation of the hidden layer's output, $h_i$, is as follows:

$$h_i = \partial(W_i x + b_i) \quad i \neq m \tag{1}$$

$$\partial(x) = max(0, x) \tag{2}$$

where *i represents the i-th hidden layer*, $\partial$ represents the activation function, $W_i$ represents the weight (also known as trainable matrix or connection coefficient), $b_i$ represents the bias and *m* is the number of hidden layers. The transition from the hidden layer to the output layer can be viewed as a multi-class logistic regression. The subsequent equations are as follows:

$$o = Softmax(W_i x + b_i) \quad i = m \tag{3}$$

$$Softmax(x)_i = \frac{e^{x_i}}{\sum_{j=1}^{n} e^{x_j}} \tag{4}$$

Where *Softmax* is the activation function of the final output, $n$ is the length of $h_i$, $x_i$ and $x_j$ is the value at the position $i$ on $h_i$. Hence, all the parameters in the MLP consist of the connection weights and biases between the layers, encompassing $W_i$ and $b_i$. When dealing with a particular problem, the gradient descent method is applied to optimize these parameters.

For a deep and wide artificial neural network, the phenomenon of overfitting often occurs due to the large number of parameters in the model. Hinton (Srivastava et al., n.d.) proposed a method called Dropout to prevent overfitting. In this method, during each training iteration, a random subset of neurons is selected, and their output values are set to zero. These selected neurons do not participate in the forward and backward propagation processes during that iteration, effectively "dropping out" these neurons temporarily. The key aspect of Dropout is that it randomly drops neurons during the training process. As a result, different neurons are dropped out in different iterations, preventing the network from relying too heavily on any single neuron. This helps reduce the complexity of the neural network and improves its generalization ability because the network is forced to adapt to various subsets during the learning process rather than becoming overly specialized in any particular portion of the data.

*2.2.2. Models for distinguishing power transformer fault types and degrees*

This paper introduces a novel approach which is not previously reported, for data input modeling based on the characteristics of the FRA curves and MLP models. Additionally, strategies for adjusting the model output dimensions are applied to accommodate various fault diagnosis tasks.

In the FRA method, each sample data obtained represents a one-dimensional vector, which signifies the magnitude of output amplitude for a particular winding in response to signal perturbations ranging from low to high frequencies. Given the MLP's ability to directly process one-dimensional vector data, and with the input layer serving as a learnable matrix capable of global positional information extraction from the vector, this paper avoids the manual feature extraction approach. Instead, it directly inputs the amplitude responses at different frequencies from the FRA curve as a one-dimensional vector into the Multilayer Perceptron model. As evidenced by the final experimental results, this approach yields a high diagnostic accuracy and generalizability.

For a Multilayer Perceptron model based on FRA data, the depth and width of the model are both subjects of exploration. In order to design an appropriate model architecture that performs well across three distinct datasets, six Multilayer Perceptron models with varying depths and widths are designed and evaluated.

**Table 5.** Details of the six MLP models

| Item | FRA-Dialight | FRA-Diagnoser | FRA-DiaL | FRA-DiaL-D | FRA-DiaXL | FRA-DiaXL-D |
|---|---|---|---|---|---|---|
| **Dropout (After)** | No | No | No | Layer 4 | No | Layer2&4 |
| **Total params** | 2M | 18M | 82M | 82M | 32M | 32M |
| **Total Size (MB)** | 9.55 | 70.73 | 315 | 315 | 122.72 | 122.76 |

**Note**: the numbers of params are calculated when set the wide of output layers as 5.

As shown in Table 5, the 6 models are named based on the depth of their learnable layers, namely, FRA-Dialight, FRA-Diagnoser, FRA-DiaL, FRA-DiaL-D, FRA-DiaXL, and FRA-DiaXL-D. And the structure of the six FRA-Dia series models were shown in Figure 3. These 6 models share the common feature of having large intermediate layer dimensions while maintaining small input and output layer dimensions. Specifically, FRA-Diagnoser is a 5-layer MLP model, FRA-DiaXL is a 10-layer MLP model, exhibiting a more profound layer hierarchy compared to the former, and FRA-DiaX is a 7-layer MLP model, which not only increases the number of layers but also expands the width of the intermediate layers, resulting in the highest number of learnable parameters. During model training, both of these larger models exhibited overfitting to the training data. Consequently, dropout layers were incorporated into these models, leading to the creation of FRA-DiaL-D and FRA-DiaXL-D. At the same time, to explore the effectiveness of lightweight models, an additional three-layer MLP model, named FRA-Dialight, was designed.

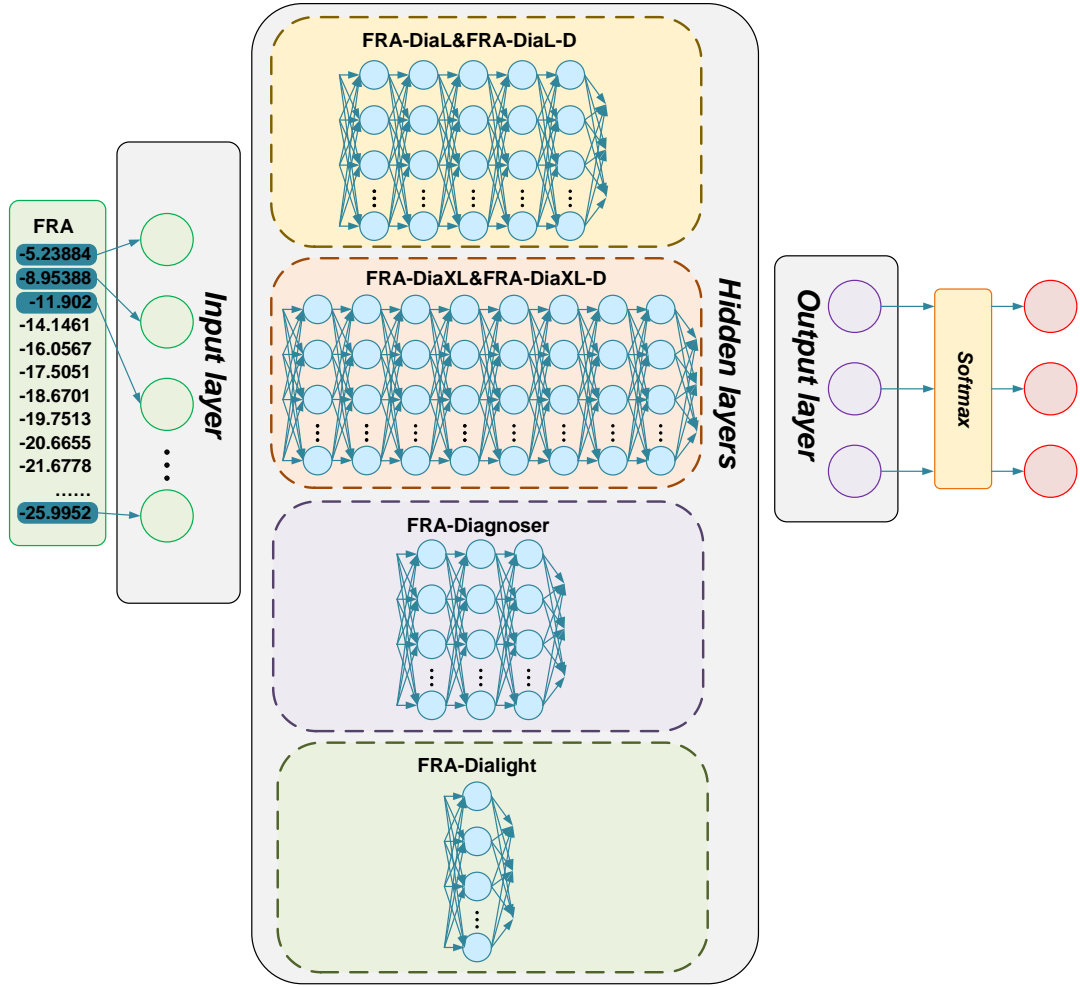

**Figure 3.** Structure of the FRA-Dia series models. The six models exhibit identical input layer dimensions and output layers configured according to different diagnostic tasks. The number of intermediate layers is set as depicted in the diagram.

For a specific model architecture, by altering the width of the learnable matrix in the final layer, the model can be adapted to different tasks across various datasets. On one hand, by training the model on fault type diagnosis tasks across different datasets and adjusting the output width to 1 (label from the normal) plus the number of fault categories, the model's capability to diagnose fault types can be evaluated. On the other hand, by training the model using data instances of the same fault type at various locations within the dataset as a label and configuring the model's output width to 1 (label from the normal) plus the number of labels corresponding to different fault severity levels, the model's diagnostic results for transformer faults are ultimately obtained through the *Softmax*. Through these two different configurations of the model's output layer, training and evaluation were conducted on six variants of the FRA-Dia series models for the diagnosis of fault occurrence and fault type, as well as the quantitative assessment of fault severity in the context of different faults.

### 2.2.3. Model training

Deep learning architecture, PyTorch, is used for model training, employing stochastic gradient descent for model training and parameter updates. The *CrossEntropyLoss* function has been chosen as the loss function:

$$CrossEntropyLoss = -\frac{1}{N}\sum_{i=1}^{N}\sum_{j=1}^{C} y_{ij} log(p_{ij}) \qquad (5)$$

Where $N$ is the number of samples. C is the number of categories. $Y_{ij}$ represents whether sample $i$ belongs to category $j$, with 1 indicating belonging and 0 indicating non-belonging. $p_{ij}$ is the probability predicted by the model that sample $i$ belongs to category $j$. Adam is chosen as the optimizer, and the learning rate has been set to 0.0001 for all the models.

### 2.2.4. Ensemble learning

Ensemble learning (Breiman, 1996) is a method that enhances classification accuracy by linearly combining the classification outcomes of multiple sub-models to create a fusion model. For an ensemble model obtained by combining two sub-models, the classification result $y$ for a sample $x$ can be represented by the following equation:

$$y = \lambda \times model_1(x) + (1-\lambda) \times model_2(x) \qquad (6)$$

Where $model_i$ represents the prediction result from the *i-th* model, and $\lambda$ is the weight parameter, which determines the model's contribution to the final prediction result.

### 2.3. Visualization of Disparities in FRA Curves Under Varied Levels of Fault Severity

In the context of FRA curves for windings at varying levels of fault severity, differences can be visualized through the utilization of numerical index. Therefore, in this paper, the Correlation Coefficient (CC) and Euclidean Distance (ED) metrics are employed for differential representation. The formulas for CC and ED are provided as follows:

$$CC(X,Y) = \frac{\sum_{i=1}^{n}(X_i - \bar{X})(Y_i - \bar{Y})}{\sqrt{\sum_{i=1}^{n}(X_i - \bar{X})^2 \sum_{i=1}^{n}(Y_i - \bar{Y})^2}} \tag{7}$$

$$ED(X,Y) = \sqrt{\sum_{i=1}^{n}(X_i - Y_i)^2} \tag{8}$$

Where $X$ and $Y$ represent the datasets of FRA curves for windings under different fault severities, $X_i$ and $Y_i$ denote individual data points within these datasets, $\bar{X}$ and $\bar{Y}$ are the means of the respective datasets, and $n$ signifies the number of data points in each dataset. In this paper, in addition to employing fault samples for the statistical visualization of the relationship charts between the CC and ED data metrics, we have also generated average curve plots. Specifically, we computed the mean FRA curve at different frequency points for all samples of the same fault type (FB fault) at the same severity level. This resultant composite curve, once visually represented, can depict the statistical distribution of a specific FB fault type.

*2.4. Evaluation metrics*

For multi-label tasks, the model performances can be evaluated through *Accuracy* (ACC) and *Macro-Averaged F1-Score* (F1). ACC can be calculated as follow:

$$ACC = \frac{TP + TN}{TP + FN + TN + FP} \tag{9}$$

Where *TP* (True Positives) represents the number of correctly predicted positive instances, *TN* (True Negatives) represents the number of correctly predicted negative instances, *FN* (False Negatives) represents the number of incorrectly predicted negative instances, and *FP* (False Positives) represents the number of incorrectly predicted positive instances. Hence, Accuracy (ACC) is the proportion of correctly classified samples to the total number of samples.

To calculate the *Macro-Averaged F1-Score* for a multi-class classification problem, *F1-Score* for each class is firstly calculated follow:

$$F1\text{-}Score = \frac{2 \times \left(\frac{TP}{TP+FP} \times \frac{TP}{TP+FN}\right)}{\frac{TP}{TP+FP} + \frac{TP}{TP+FN}} \quad (10)$$

And then, divide the sum of *F1-Score* for each class by the total number of classes (*C*) to get the *Macro-Averaged F1-Score*:

$$Macro\text{-}Averaged\ F1\text{-}Score = \frac{\sum F1\text{-}Scores\ for\ all\ classes}{C} \quad (11)$$

Thus, the Macro-Averaged F1-Score offers a means to compute the mean F1-Score across all class categories, affording equitable consideration to each class.

In order to accurately assess the performance of the models and to fairly compare the diagnostic efficacy of different models, a 10-fold cross-validation approach is employed in this study. The principle of 10-fold cross-validation involves partitioning the dataset into ten subsets, or folds, with one-fold utilized as the validation set while the remaining nine folds are employed for training the model. This process is iteratively repeated ten times, each time using a different fold as the validation set. The results from each iteration are then averaged to obtain a comprehensive evaluation of the model's performance, thereby mitigating the impact of dataset variability and enhancing the reliability of the findings. This rigorous methodology is employed to ensure robustness and objectivity in the assessment of the models under consideration.

## 3. Results and discussion

### 3.1. Model performances on different datasets

*3.1.1. Model performances on fault type diagnose*

The study conducted training and evaluation of six FRA-Dia series models primarily for the task of diagnosing various types of faults in power transformers. The results of a 10-fold cross-validation are presented in Table 6. Across three datasets, FRA-Diagnoser achieved ACC of 100%, 99.7%, and 99.8%, respectively, which were the highest among all models. Furthermore, it exhibited the lowest standard errors among the six Multilayer perceptron models. The F1 for FRA-Diagnoser were 100%, 97.4%, and 97.9%, also the highest among the models. These findings indicate that FRA-Diagnoser displayed the highest precision in diagnosing different types of faults and, therefore, stands as the most potent diagnostic model for transformer winding fault categorization.

In the case of FRA-DiaL and FRA-DiaXL, both models achieved ACC rates exceeding 80%. However, their overall performance was inferior to that of FRA-Diagnoser, suggesting potential overfitting in these two models. Even with the incorporation of Dropout layers in the larger models, their performance could not surpass that of FRA-Diagnoser. The underlying reason for this phenomenon lies in the limited number of label categories in the task of classifying power transformer fault types. With a fixed amount of data samples, larger models become more susceptible to overfitting as training iterations increase.

Notably, FRA-Dialight achieved ACC rates of 99.2%, 98.2%, and 99.8% across the three datasets, showing only a slight decrease in performance compared to FRA-Diagnoser. Moreover, it exhibited minimal standard errors (0.02, 0.02, and 0.01), which were the smallest among the models excluding FRA-Diagnoser. However, model size of FRA-Dialight is only one-ninth of that of FRA-Diagnoser. Additionally, it boasted faster model training, deployment inference speed, and less than one-seventh of the model memory footprint.

Hence, FRA-Dialight holds substantial practical value in the field of on-site power transformer inspection. Particularly, in scenarios where the sole requirement is to diagnose power transformer winding fault types, FRA-Dialight can be deployed as a diagnostic model. This represents a swift, efficient, and cost-effective solution that can run even on older CPU-based host systems.

Table 6. Model performances on fault type diagnose.

| Dataset | Model | ACC | F1 |
|---|---|---|---|
| Group1 (EE winding 10-disc) | FRA-Dialight | 0.992±0.02 | 0.994±0.01 |
|  | **FRA-Diagnoser** | **1.000±0.00** | **1.000±0.00** |
|  | FRA-DiaL | 0.933±0.09 | 0.845±0.23 |
|  | FRA-DiaL-D | 0.926±0.09 | 0.803±0.25 |
|  | FRA-DiaXL | 0.911±0.15 | 0.830±0.22 |
|  | FRA-DiaXL-D | 0.800±0.20 | 0.674±0.25 |
| Group2 (CIW winding 10-disc) | FRA-Dialight | 0.982±0.02 | 0.885±0.14 |
|  | **FRA-Diagnoser** | **0.997±0.01** | **0.974±0.08** |
|  | FRA-DiaL | 0.835±0.24 | 0.68±0.282 |
|  | FRA-DiaL-D | 0.914±0.18 | 0.823±0.26 |
|  | FRA-DiaXL | 0.879±0.17 | 0.711±0.28 |
|  | FRA-DiaXL-D | 0.873±0.12 | 0.693±0.23 |
| Group3 (EE winding 12-disc) | FRA-Dialight | 0.998±0.01 | 0.979±0.07 |
|  | **FRA-Diagnoser** | **0.998±0.00** | **0.979±0.06** |
|  | FRA-DiaL | 0.998±0.01 | 0.979±0.07 |
|  | FRA-DiaL-D | 0.998±0.01 | 0.979±0.07 |
|  | FRA-DiaXL | 0.938±0.09 | 0.783±0.19 |
|  | FRA-DiaXL-D | 0.901±0.15 | 0.762±0.20 |

**Note:** mean±standard error from 10-fold cross-validation, rounding to the nearest integer.

In summary, the FRA-Diagnoser model exhibited the most superior performance across all datasets, indicating that it can accurately determine the presence of faults in power transformers through FRA curve analysis. Moreover, it can precisely identify the specific fault types present in the transformers. Importantly, the diagnostic performance of this model is not restricted by the number of windings in the transformer or the measurement wiring configurations used in the FRA data.

### 3.1.2. Model performances on fault degree diagnose

The diagnostic performance of the six perceptron models for assessing the severity of power transformer faults is presented in Table 7.

Table 7. Model performances on fault degree diagnose.

| Dataset | Model | ACC | F1 |
| --- | --- | --- | --- |
| **Group1** (EE winding 10-disc) | **FRA-Dialight** | **0.902±0.07** | **0.903±0.06** |
| | FRA-Diagnoser | 0.869±0.10 | 0.864±0.11 |
| | FRA-DiaL | 0.646±0.16 | 0.597±0.17 |
| | FRA-DiaL-D | 0.412±0.14 | 0.337±0.17 |
| | FRA-DiaXL | 0.634±0.27 | 0.597±0.31 |
| | FRA-DiaXL-D | 0.538±0.22 | 0.472±0.27 |
| **Group2** (CIW winding 10-disc) | FRA-Dialight | 0.972±0.04 | 0.934±0.07 |
| | **FRA-Diagnoser** | **0.996±0.01** | **0.983±0.05** |
| | FRA-DiaL | 0.735±0.28 | 0.676±0.33 |
| | FRA-DiaL-D | 0.752±0.22 | 0.705±0.27 |
| | FRA-DiaXL | 0.701±0.13 | 0.599±0.19 |
| | FRA-DiaXL-D | 0.727±0.16 | 0.680±0.20 |
| **Group3** (EE winding 12-disc) | **FRA-Dialight** | **0.925±0.04** | **0.906±0.05** |
| | FRA-Diagnoser | 0.875±0.16 | 0.836±0.21 |
| | FRA-DiaL | 0.732±0.19 | 0.654±0.24 |
| | FRA-DiaL-D | 0.742±0.19 | 0.701±0.23 |
| | FRA-DiaXL | 0.640±0.14 | 0.479±0.16 |
| | FRA-DiaXL-D | 0.767±0.10 | 0.688±0.14 |

**Note:** mean±standard error from 10-fold cross-validation, rounding to the nearest integer.

For two datasets of transformer winding FRA data obtained using the EE connection method, FRA-Dialight consistently demonstrated the best performance. It achieved ACC metrics exceeding FRA-Diagnoser by 3% and 5%, and F1 scores surpassing FRA-Diagnoser by 4% and 7%. Furthermore, FRA-Dialight exhibited the lowest diagnostic standard errors among all the models. Therefore, FRA-Dialight, when applied to the EE connection method, not only accurately determines the presence of winding faults and their specific types but also provides the most precise diagnosis of fault severity. In contrast, the other five models, which have a higher number of learnable parameters, exhibited relatively poorer diagnostic accuracy for transformer frequency response data acquired through the EE connection, suggesting that these models may have experienced varying degrees of overfitting. Consequently, these five models are not the optimal choice for winding fault detection under EE connection conditions.

For the transformer winding FRA data acquired using the CIW connection method,

FRA-Diagnoser displayed the best performance, with ACC and F1 metrics reaching 99.6% and 98.3%, respectively. It also exhibited the smallest standard errors, signifying outstanding diagnostic capabilities across various types of power transformer fault severities. Compared to the best model, FRA-Dialight, trained using the EE connection method for a 10-disc winding power transformer, FRA-Diagnoser outperformed by more than 9% and 8% on ACC and F1. Therefore, utilizing the CIW connection method for FRA data acquisition in conjunction with FRA-Diagnoser for diagnostics appears to be the most effective approach for assessing the severity of transformer faults, as per the results presented in this study.

### *3.2. Comparative analysis model performances and model utilization strategies*

### *3.2.1. Model application strategies for transformer winding fault detection and fault type determination*

Different models exhibit varying diagnostic performance for different types of faults. In order to further analyze and compare the performance of these six models in diagnosing faults in power transformers, we employed a 10-fold cross-validation approach to generate confusion matrices based on diagnostic results and the ground truth labels of the samples. The results in these confusion matrices provide an intuitive assessment of the models' performance in diagnosing both the fault type and the severity of a specific fault type across the entire dataset.

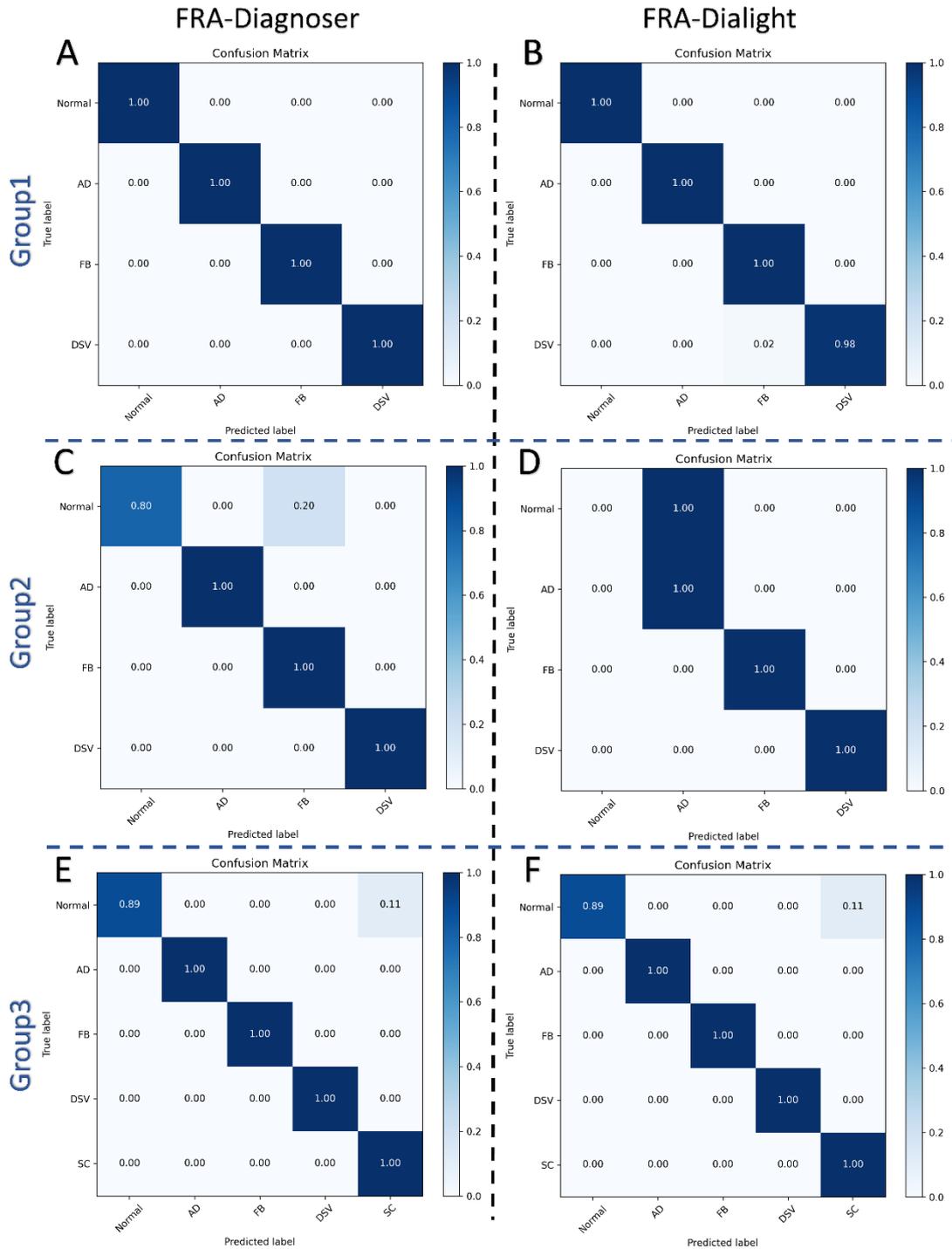

**Figure 4.** The confusion matrixes from FRA-Diagnoser and FRA-Dialight on different dataset for diagnosing the fault type.

From the perspective of ACC and the F1, it is evident that FRA-Diagnoser and FRA-Dialight demonstrate the highest performance. The confusion matrices representing their diagnostic results for different fault types are depicted in Figure 4. It is noteworthy that both FRA-Diagnoser and FRA-Dialight exhibit exceptionally high accuracy in diagnosing whether Group1 (comprising FRA data acquired using the EE wiring configuration from a 10-disc winding transformer) is faulty and in distinguishing between fault types. However, for Group2 (consisting of FRA data obtained using the CIW wiring configuration from a 10-disc winding transformer), these models can only achieve 100% accurate differentiation among fault samples, while occasionally misidentifying normal samples. Notably, FRA-Dialight tends to misclassify all normal transformer windings as windings with AD faults. This misclassification is primarily attributed to the relatively shallow depth of the FRA-Dialight model and the comparatively narrower width of each layer, which causes it to perceive normal power transformer FRA data acquired using the CIW wiring configuration as being similar to certain locations with mild AD faults, as illustrated in Figure 5. (B).

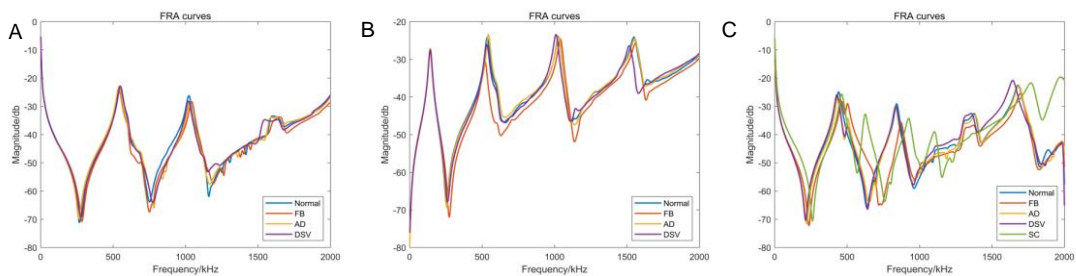

**Figure 5.** The FRA curve of the transformer from the Dataset Group1 (A), Group2 (B) and Group3 (C), as well as the first fault sample with the mildest fault severity among three different fault types. This image was produced through the visualization of the FRA data.

Within the dataset obtained using the CIW wiring configuration, there is a distinct overlap between samples representing normal operation and those with the mildest degree of AD faults, as shown in Figure 5. (A) and Figure 5. (C) In contrast, significant differences exist between samples of normal transformer windings and those with AD faults when using the EE wiring configuration, as illustrated in Figure 5. (C). Therefore, for the specific diagnostic problem of AD faults, FRA data acquired using the CIW wiring configuration yields higher diagnostic accuracy. In the case of Group3 dataset obtained using the EE wiring configuration, FRA data from transformers with mild SC faults bears a strong resemblance to the FRA data from normal power transformers. Consequently, the models occasionally misclassify normal samples as fault samples.

In conclusion, when the presence of a fault is already known, utilizing either the FRA-Dialight or FRA-Diagnoser models in isolation for fault type diagnosis results in a 100% accuracy rate. However, in scenarios where both the determination of whether a transformer is faulty and the diagnosis of the specific fault type are required, the integration of FRA data acquired through the EE method enhances the diagnostic performance of these Multilayer Perceptron models.

### *3.2.2. Model application strategies for transformer winding fault severity diagnosis requirements*

The performances of the models in fault severity discrimination are depicted in Figure 6, while the RFA curves of a power transformer winding exhibiting varying degrees of FB faults are shown in Figure 7. (A, C, E). Furthermore, the curves were subjected to statistical visualization through the employment of both the CC and ED metrics, yielding Figure 7. (B, D, F).

**Figure 6.** The confusion matrixes from FRA-Diagnoser and FRA-Dialight on different dataset for diagnosing the fault type with fault degree.

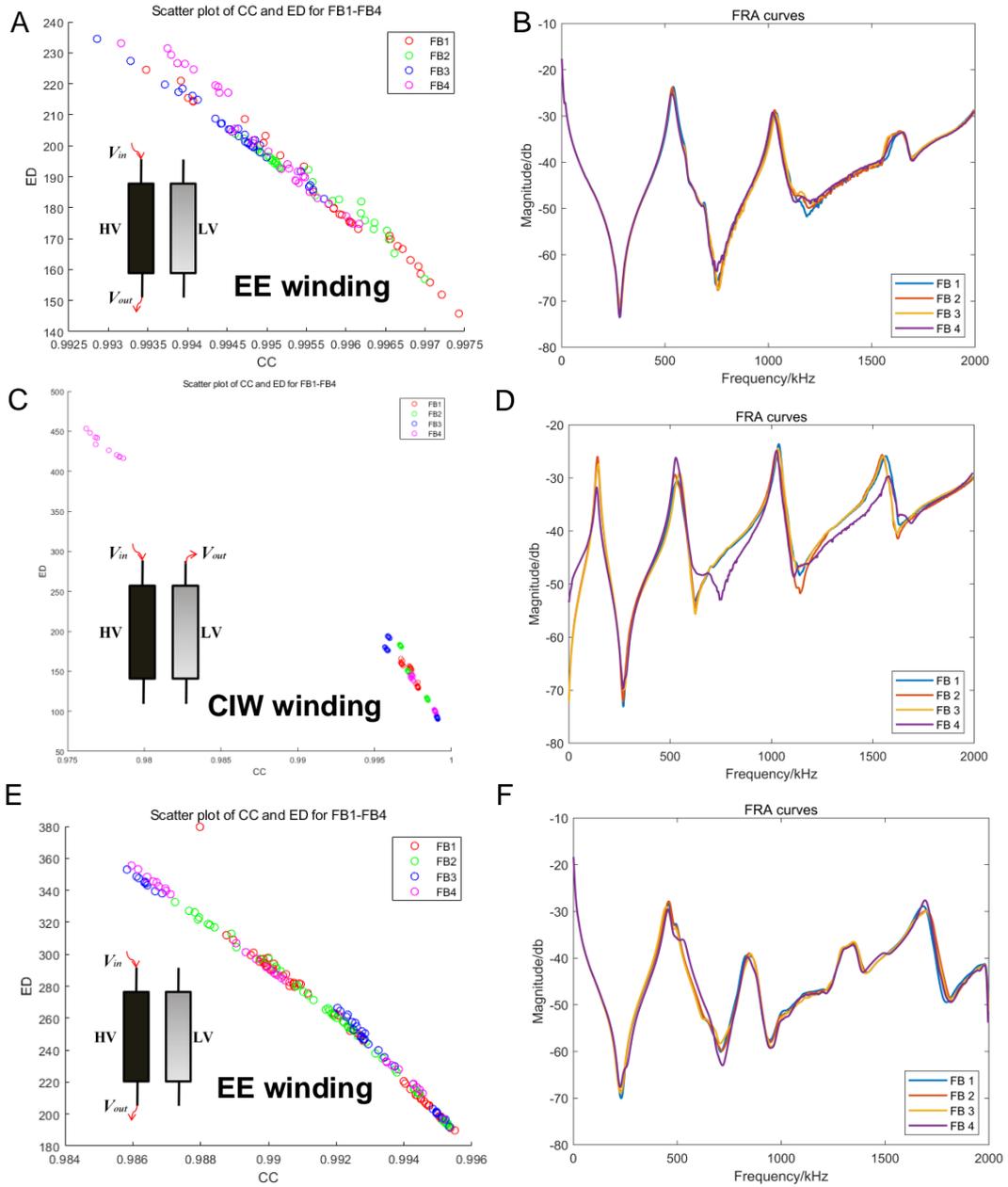

**Figure 7.** Samples of Group1 (A, B), Group2 (C, D), Group3 (E, F). (A, C, E) The curve represents the mean values of all samples measured at various locations and times within a specific fault severity level in the given dataset. (B, D, F) The statistical visualization of CC and ED metrics (CC-ED Map).

Due to the limited differentiation in the RFA curves obtained from transformers with four different degrees of FB faults using the EE wiring configuration, as illustrated in Figure 6. (A, B, E, F), various MLPs exhibit a certain degree of diagnostic errors. Simultaneously, from the relationship figures of CC and ED, it is evident that there exist significant statistical variances in the samples obtained under EE connections for each level of fault severity. This poses challenges for the model in capturing statistical information, thus leading to diagnostic errors in severity assessment. In comparison to FRA-Diagnoser, FRA-Dialight, owing to its fewer model parameters and the absence of substantial feature extraction capabilities for deep FRA data, shows more pronounced diagnostic inaccuracies, particularly in the case of FB faults with degrees 2 and 3. Notably, for data acquired using the CIW wiring configuration, there are significant differences in the RFA curves for FB faults with different degrees. The statistical characteristics of FRA data obtained under the CIW connection configuration are depicted in Figure 7. (C, D) Whether one examines the average FRA curves for the four fault severity levels or inspects the relationship charts of CC and ED, it is evident that there exist substantial statistical disparities among the curves under different fault severity levels. The symmetrical winding structure of EE windings, as observed from the input-output voltage relationship, leads to challenges in fault diagnosis using FRA. Specifically, symmetrical faults such as those occurring on the first and last disks of the high-voltage winding side exhibit similar FRA signatures. Consequently, as depicted in Figure 7. (A, E), the sample distribution from the CC-ED Map forms a linear pattern. All above is reflected in the model's diagnostic performance, where confusion arises in classifying the four distinct degrees of FB. Conversely, the asymmetrical nature of the CIW connection facilitates effective fault diagnosis, as it results in more discernible FRA signatures for different fault types and degrees. As a result, FRA-Diagnoser achieves a 100% accuracy in diagnosing the severity of FB faults, with model diagnostic errors primarily occurring in the detection of normal samples. Consequently, it is recommended to employ the CIW wiring method for FRA winding tests in situations where the power transformer fault is known, focusing on the detection of fault types and severity based on the corresponding FRA data. In summary, for the diagnosis of fault severity in various fault types, both FRA-Diagnoser and FRA-Dialight yield favorable diagnostic results.

## 3.3. Model comparison

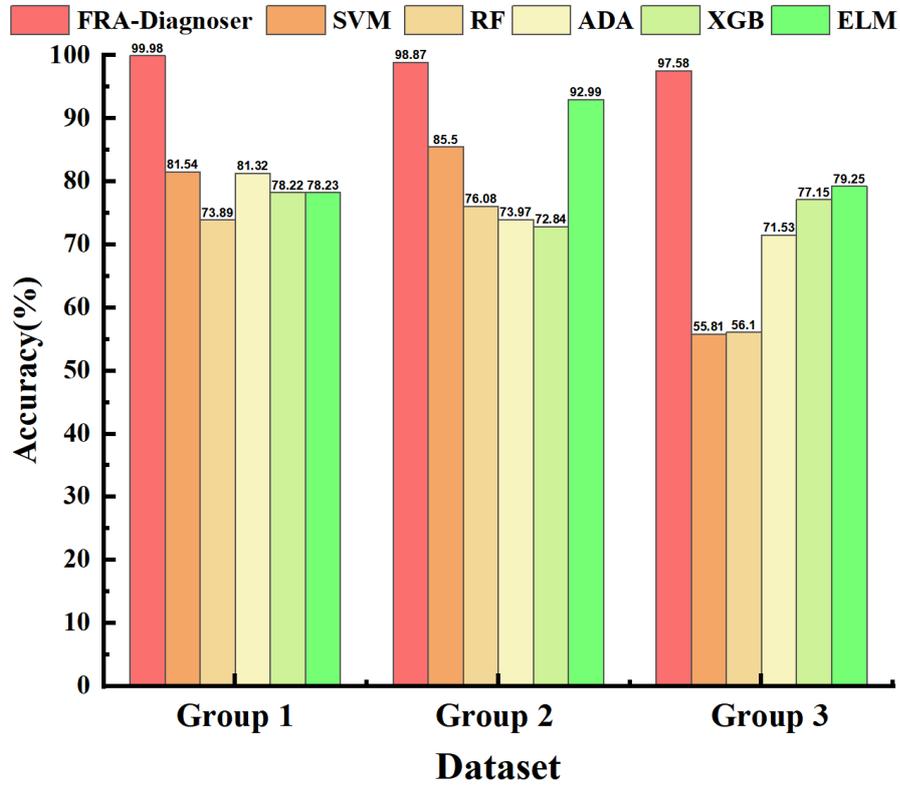

**Figure 8.** Model comparison on 10-fold cross validation.

The models were evaluated and compared using a 10-fold cross-validation approach, with the fault type diagnosis results are presented in Figure 8. Here, all the models are separately trained on evaluated on the 3 datasets. SVM and RF which has been introduced into the winding fault diagnosis with numeral index input were included in the comparison. Additionally, AdaBoost (ADA), XGBoost (XGB), and Extreme Learning Machine (ELM) were also incorporated with the same data input formation (Pedregosa et al., n.d.). Among all the six models, FRA-Diagnoser achieved the best performance on all the three datasets, surpassing all the models by more than 0.18, 0.05, and 0.18 on data Group 1 to Group 3. Therefore, FRA-Diagnoser is the best model for winding fault detection and fault type diagnosis with generalizability and robustness.

## 3.4. Model utilization strategy for the two-stage approach to transformer winding fault diagnosis

In the pursuit of further enhancing the performance of transformer winding fault diagnosis models based on the FRA method and achieving the ultimate models that excel in various problem domains, we employed an ensemble learning approach. However, the fusion model derived from this approach failed to yield satisfactory results. Consequently, we propose a two-stage diagnostic strategy based on multiple models.

### 3.4.1. Performance evaluation of the fusion model

A comparative analysis of the performance of the six models across the three datasets reveals that FRA-Diagnoser and FRA-Dialight exhibit the best performance. Moreover, these two models demonstrate distinct performance advantages in different fault diagnosis tasks. Therefore, theoretically, combining these two models through ensemble learning can yield a comprehensive performance surpassing that of each individual model. By adjusting the weight parameters during the fusion process, the final fusion model's results for transformer winding fault detection are shown in Table 8.

**Table 8.** Diagnose performance of the fusion model.

| Task | Dataset | ACC | F1 |
|---|---|---|---|
| **Fault type** | EE 10 | 0.993±0.02 | 0.995±0.02 |
| | CIW 10 | 0.982±0.03 | 0.885±0.15 |
| | EE 12 | 0.998±0.01 | 0.979±0.07 |
| **Fault type** | EE 10 | 0.902±0.07 | 0.903±0.06 |
| **(degree)** | CIW 10 | 0.921±0.04 | 0.934±0.07 |
| | **EE 12** | **0.925±0.04** | **0.906±0.05** |

**Note:** mean±standard error from 10-fold cross-validation, rounding to the nearest integer.

In comparison to the results of fault diagnosis using FRA-Dialight or FRA-Diagnoser individually, the fusion model achieves fault diagnosis performance equivalent to that of FRA-Dialight only in the case of fault severity discrimination using FRA data obtained from the 12-disc winding transformer windings with the EE wiring configuration. Traditionally, it has been assumed that fusion models often outperform all their constituent sub-models. However, the results in Table 8 indicate that these fusion models mostly exhibit weaker diagnostic performance compared to the better-performing sub-models for most of the scenarios.

Two main reasons contribute to this phenomenon. Firstly, both sub-models have high accuracy rates across all tasks, particularly in the task of fault type detection, with FRA-Diagnoser achieving accuracy rates exceeding 99%. Therefore, the utility of ensemble learning methods in improving the diagnostic performance of individual models is limited. Moreover, the performance of FRA-Dialight, slightly inferior by itself, could lead to a decrease in the final diagnostic accuracy.

On the other hand, the distribution of misclassified samples for the two sub-models differs due to their varying sensitivities to diagnosing different winding faults. As illustrated in Figure 6. (D), the samples where the sub-models make classification errors are highly imbalanced, with errors predominantly concentrated in a single category. Consequently, even if one sub-model exhibits a good recognition accuracy for such samples, it can be severely disrupted by the other sub-model that commits errors, resulting in a degradation in the performance of the final fusion model.

In summary, ensemble learning methods exhibited suboptimal performance for the fault diagnosis problem of power transformer windings considered in this study. Consequently, we propose a two-stage model utilization strategy based on the FRA-Diagnoser and FRA-Dialight models.

### *3.4.2. Two-stage model utilization strategy for transformer winding fault diagnosis*

Based on the experimental results presented above, it is evident that, with regard to the issue of detecting faults in transformer windings, the FRA data obtained using different models in combination with different connection methods exhibit varying performance in diagnosing different types of faults. Consequently, rather than following a parallel design approach like ensemble learning, a serial model utilization strategy is proposed, which builds upon the diagnostic output of one perceptron model to be used as the basis for another perceptron model. The fundamental reason behind proposing this strategy is the different sensitivities of different connection methods to various winding fault types, resulting in either strong or weak overlaps in the FRA curves for different fault scenarios. This overlap can interfere with the model's decision-making process, leading to erroneous diagnostic results. From a statistical perspective, conducting FRA tests on the same winding with several different connection methods can be understood as tools for feature extraction from the physical structure of the winding along different dimensions. For example, EE and CIW connections produce different FRA curves for the same winding fault, indicating that these two methods focus on different dimensions for the same fault scenario, and therefore provide different 'conclusions' for the same problem. By training multiple models separately using features extracted from EE and CIW connection methods, the serial model utilization strategy leverages these 'conclusions' from multiple dimensions as criteria for the final diagnosis of winding faults. Therefore, the two-stage model utilization strategy offers better performance in diagnosing winding faults. On the other hand, the diagnosis of specific fault conditions is a progressive process, involving first determining whether the winding is faulty, then identifying the specific fault type, and finally assessing the degree of winding fault. Therefore, the serial two-stage model utilization strategy outperforms the parallel decision-making ensemble learning method in the quantitative diagnosis of transformer winding fault types and degrees. From the results in Figure 8, it can be seen that by using the two-stage model, a 100% accurate quantitative diagnosis of transformer winding fault types and degrees for the 10-disc transformer discussed in this paper is achieved.

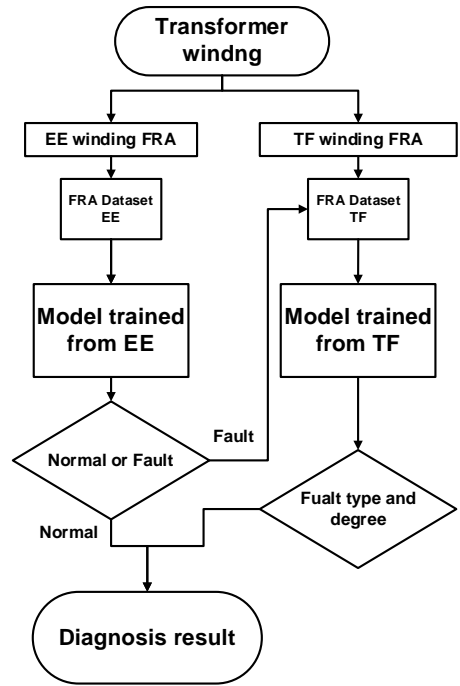

**Figure 9.** The two-stage method for using transformer fault diagnosis models.

As shown in Figure 9, for a specific model of power transformer, a two-stage transformer winding diagnosis model can be developed by training multiple models using FRA data sets obtained under different connection methods (especially common ones like EE and CIW). The order of model utilization is determined based on the diagnostic performance advantages of different models. For example, in the case of fault detection in a 10-disc transformer winding, the FRA-Diagnoser trained with FRA data obtained through EE connection can first diagnose whether there is a fault in the transformer winding. If the diagnosis result indicates a fault, the FRA-Diagnoser trained with FRA data obtained through CIW connection can then be used to perform quantitative diagnosis of the fault type and degree. This two-stage model utilization approach effectively leverages the strengths of different models to obtain accurate quantitative diagnostic results.

Additionally, in certain specific cases, only the second-stage model can be used for the final fault diagnosis. For instance, in the case of a faulty power transformer, FRA testing using the CIW connection method can be conducted before dismantling and maintenance. The obtained data can be feed into a pre-trained FRA-Diagnoser or FRA-Dialight model, and the model's diagnostic results can be used for targeted assessment and repair of specific faults in the transformer winding.

*3.5. Implementation and limitation*

From the model evaluation results presented in this paper, it is evident that models such as the FRA-Diagnoser, based on FRA data, exhibit exceptional performance in the quantitative diagnosis of transformer winding faults. Furthermore, by employing a two-stage approach, a 100% accuracy in fault diagnosis performance is achievable for the 10-disc winding power transformer discussed in this study. Due to the limited availability of FRA data for transformer windings, especially for those with identified fault labels, the FRA data is even scarcer. Therefore, this article exclusively demonstrates the diagnostic performance of FRA-Diagnoser and FRA-Dialight models with MLP architecture on three distinct sets of simulated transformer winding FRA data. It should be noted, however, that these three FRA data sets in the comparative analysis encompass various transformer winding models of different types, FRA measurements obtained under different connection configurations, and within each dataset, they encompass normal samples under varying operational conditions and fault samples of varying severities and locations. Consequently, the diagnostic methodology proposed in this paper for transformer winding FRA data is characterized by its generality. Furthermore, the models presented in this paper require further validation with a more diverse and comprehensive collection of transformer winding FRA data from various sources.

In real-world scenarios involving power transformers, the working conditions are often more complex, with higher voltage levels and a greater number of transformer discs. Consequently, FRA curves obtained under faulty conditions provide a wealth of information for learning. Therefore, the large-scale models like FRA-DiaXL utilized in this study may offer even better diagnostic performance. Furthermore, by employing the two-stage model utilization strategy proposed in this paper, it is possible to create a high-precision expert diagnostic system for transformers operating in real-world scenarios using FRA data collected with EE and CIW connections.

## 4. Conclusion

In this paper, MLP-based models directly using FRA data from transformer windings are proposed. Based on this approach, six sets of varying-sized model architectures have been designed. The performance of these models on different transformer winding diagnostic tasks was evaluated by creating three sets of simulated windings. Favorable performance was exhibited by the FRA-Diagnoser and FRA-Dialight models. Furthermore, a two-stage model utilization method has been introduced, which utilizes information present in the FRA curves obtained from different connection methods used in FRA testing for the transformer windings under investigation. As a result, a transformer winding fault diagnosis system has been developed that achieves 100% accuracy on the data from a winding-10 transformer presented in this paper.

## CRediT authorship contribution statement

**Guohao Wang**: Formal analysis, Conceptualization, Investigation, Methodology, Visualization, Writing - original draft.

## Declaration of competing interest

The authors declare that they have no known competing financial interests or personal relationships that could have appeared to influence the work reported in this paper.

## References

Ahila, R., Sadasivam, V., & Manimala, K. (2015). An integrated PSO for parameter determination and feature selection of ELM and its application in classification of power system disturbances. *Applied Soft Computing, 32*, 23–37. https://doi.org/10.1016/j.asoc.2015.03.036

Akhavanhejazi, M., Gharehpetian, G. B., Faraji-dana, R., Moradi, G. R., Mohammadi, M., & Alehoseini, H. A. (2011). A new on-line monitoring method of transformer winding axial displacement based on measurement of scattering parameters and decision tree. *Expert Systems with Applications*, *38*(7), 8886–8893. https://doi.org/10.1016/j.eswa.2011.01.100

Behjat, V., & Mahvi, M. (2015). Statistical approach for interpretation of power transformers frequency response analysis results. *IET Science, Measurement & Technology*, *9*(3), 367–375. https://doi.org/10.1049/iet-smt.2014.0097

Bigdeli, M., & Abu-Siada, A. (2022). Clustering of transformer condition using frequency response analysis based on k-means and GOA. *Electric Power Systems Research*, *202*, 107619. https://doi.org/10.1016/j.epsr.2021.107619

Bigdeli, M., Siano, P., & Alhelou, H. H. (2021). Intelligent Classifiers in Distinguishing Transformer Faults Using Frequency Response Analysis. *IEEE Access*, *9*, 13981–13991. https://doi.org/10.1109/ACCESS.2021.3052144

Bigdeli, M., Vakilian, M., & Rahimpour, E. (2012). Transformer winding faults classification based on transfer function analysis by support vector machine. *IET Electric Power Applications*, *6*(5), 268–276. https://doi.org/10.1049/iet-epa.2011.0232

Breiman, L. (1996). Bagging predictors. *Machine Learning*, *24*(2), 123–140. https://doi.org/10.1007/BF00058655

Duan, J., He, Y., & Wu, X. (2021). Serial transfer learning (STL) theory for processing data insufficiency: Fault diagnosis of transformer windings. *International Journal of Electrical Power & Energy Systems*, *130*, 106965. https://doi.org/10.1016/j.ijepes.2021.106965

He, K., Zhang, X., Ren, S., & Sun, J. (2016). Deep Residual Learning for Image Recognition. *2016 IEEE Conference on Computer Vision and Pattern Recognition (CVPR)*, 770–778. https://doi.org/10.1109/CVPR.2016.90

Huang, G.-B., Zhu, Q.-Y., & Siew, C.-K. (2006). Extreme learning machine: Theory and applications. *Neurocomputing*, *70*(1–3), 489–501. https://doi.org/10.1016/j.neucom.2005.12.126

*IEEE SA - IEEE C57.149-2012*. (n.d.). Retrieved October 29, 2023, from https://standards.ieee.org/ieee/C57.149/5216/

Krizhevsky, A., Sutskever, I., & Hinton, G. E. (2012). ImageNet Classification with Deep Convolutional Neural Networks. *Advances in Neural Information Processing Systems*, *25*. https://proceedings.neurips.cc/paper/2012/hash/c399862d3b9d6b76c8436e924a68c45b-Abstract.html

LeCun, Y., Bengio, Y., & Hinton, G. (2015). Deep learning. *Nature*, *521*(7553), 436–444. https://doi.org/10.1038/nature14539

Ludwikowski, K., Siodla, K., & Ziomek, W. (2012). Investigation of transformer model winding deformation using sweep frequency response analysis. *IEEE Transactions on Dielectrics and Electrical Insulation*, *19*(6), 1957–1961.

https://doi.org/10.1109/TDEI.2012.6396953

Mao, X., Wang, Z., Crossley, P., Jarman, P., Fieldsend-Roxborough, A., & Wilson, G. (2020). Transformer winding type recognition based on FRA data and a support vector machine model. *High Voltage*, *5*(6), 704–715. https://doi.org/10.1049/hve.2019.0294

Mitchell, S. D., & Welsh, J. S. (2017). Methodology to locate and quantify radial winding deformation in power transformers. *High Voltage*, *2*(1), 17–24. https://doi.org/10.1049/hve.2016.0085

Moradzadeh, A., Moayyed, H., Mohammadi-Ivatloo, B., Gharehpetian, G. B., & Aguiar, A. P. (2022). Turn-to-Turn Short Circuit Fault Localization in Transformer Winding via Image Processing and Deep Learning Method. *IEEE Transactions on Industrial Informatics*, *18*(7), 4417–4426. https://doi.org/10.1109/TII.2021.3105932

Parkash, C., & Abbasi, A. R. (2023). Transformer's frequency response analysis results interpretation using a novel cross entropy based methodology. *Scientific Reports*, *13*(1), Article 1. https://doi.org/10.1038/s41598-023-33606-0

Pedregosa, F., Varoquaux, G., Gramfort, A., Michel, V., Thirion, B., Grisel, O., Blondel, M., Prettenhofer, P., Weiss, R., Dubourg, V., Vanderplas, J., Passos, A., & Cournapeau, D. (n.d.). Scikit-learn: Machine Learning in Python. *MACHINE LEARNING IN PYTHON*.

Picher, P., Tenbohlen, S., Lachman, M., Scardazzi, A., & Patel, P. (2017). Current state

of transformer FRA interpretation. *Procedia Engineering*, *202*, 3–12. https://doi.org/10.1016/j.proeng.2017.09.689

Pourhossein, K., Gharehpetian, G. B., Rahimpour, E., & Araabi, B. N. (2012). A probabilistic feature to determine type and extent of winding mechanical defects in power transformers. *Electric Power Systems Research*, *82*(1), 1–10. https://doi.org/10.1016/j.epsr.2011.08.010

Rumelhart, D. E., Hinton, G. E., & Williams, R. J. (1986). Learning representations by back-propagating errors. *Nature*, *323*(6088), Article 6088. https://doi.org/10.1038/323533a0

Samimi, M. H., & Tenbohlen, S. (2017). FRA interpretation using numerical indices: State-of-the-art. *International Journal of Electrical Power & Energy Systems*, *89*, 115–125. https://doi.org/10.1016/j.ijepes.2017.01.014

Secue, J. R., & Mombello, E. (2008). Sweep frequency response analysis (SFRA) for the assessment of winding displacements and deformation in power transformers. *Electric Power Systems Research*, *78*(6), 1119–1128. https://doi.org/10.1016/j.epsr.2007.08.005

Srivastava, N., Hinton, G., Krizhevsky, A., Sutskever, I., & Salakhutdinov, R. (n.d.). *Dropout: A Simple Way to Prevent Neural Networks from Overfitting*.

Sutskever, I., Vinyals, O., & Le, Q. V. (2014). Sequence to sequence learning with neural networks. *Proceedings of the 27th International Conference on Neural Information Processing Systems - Volume 2*, 3104–3112.

Tahir, M., & Tenbohlen, S. (2021). Transformer Winding Condition Assessment Using Feedforward Artificial Neural Network and Frequency Response Measurements. *Energies*, *14*(11), 3227. https://doi.org/10.3390/en14113227

Vaswani, A., Shazeer, N., Parmar, N., Uszkoreit, J., Jones, L., Gomez, A. N., Kaiser, Ł., & Polosukhin, I. (2017). Attention is All you Need. *Advances in Neural Information Processing Systems*, *30*. https://proceedings.neurips.cc/paper_files/paper/2017/hash/3f5ee243547dee91fbd053c1c4a845aa-Abstract.html

Wang, S., Wang, S., Feng, H., Guo, Z., Wang, S., & Li, H. (2018). A New Interpretation of FRA Results by Sensitivity Analysis Method of Two FRA Measurement Connection Ways. *IEEE Transactions on Magnetics*, *54*(3), 1–4. https://doi.org/10.1109/TMAG.2017.2743986

Wong, P. K., Yang, Z., Vong, C. M., & Zhong, J. (2014). Real-time fault diagnosis for gas turbine generator systems using extreme learning machine. *Neurocomputing*, *128*, 249–257. https://doi.org/10.1016/j.neucom.2013.03.059

Zhao, Z., Tang, C., Chen, Y., Zhou, Q., Yao, C., & Islam, S. (2021). Interpretation of transformer winding deformation fault by the spectral clustering of FRA signature. *International Journal of Electrical Power & Energy Systems*, *130*, 106933. https://doi.org/10.1016/j.ijepes.2021.106933

Zhao, Z., Tang, C., Zhou, Q., Xu, L., Gui, Y., & Yao, C. (2017). Identification of Power Transformer Winding Mechanical Fault Types Based on Online IFRA by

Support Vector Machine. *Energies*, *10*(12), Article 12. https://doi.org/10.3390/en10122022

Zhao, Z., Yao, C., Tang, C., Li, C., Yan, F., & Islam, S. (2019). Diagnosing Transformer Winding Deformation Faults Based on the Analysis of Binary Image Obtained From FRA Signature. *IEEE Access*, *7*, 40463–40474. https://doi.org/10.1109/ACCESS.2019.2907648

Zhou, L., Jiang, J., Zhou, X., Wu, Z., Lin, T., & Wang, D. (2020). Detection of transformer winding faults using FRA and image features. *IET Electric Power Applications*, *14*(6), 972–980. https://doi.org/10.1049/iet-epa.2019.0933